\documentclass[aps,pra, footinbib,twocolumn,showpacs,amsmath,amssymb,superscriptaddress]{revtex4-2}
\usepackage{graphicx}
\usepackage{dcolumn}
\usepackage{bm}
\usepackage{hyperref}
\usepackage{multirow} 
\usepackage {color} 
\usepackage{amsmath, amssymb}
\usepackage{array}

\hypersetup{colorlinks=true, citecolor=blue, urlcolor=blue, linkcolor=blue}

\begin{document}

\title{Non-Hermiticity of an anomalous superradiant phase}
 \author{You-Qi Lu}
 \thanks{These two authors contributed equally}
 \affiliation{Department of Physics, Chongqing Key Laboratory for strongly coupled Physics,
Chongqing University, Chongqing 401330, China}
\author{Jun-Ling Wang}
 \thanks{These two authors contributed equally}
 \affiliation{Zhejiang Key Laboratory of Micro-Nano Quantum Chips and Quantum Control, School of Physics, Zhejiang University, Hangzhou 310027, China}
\author{Qing-Hu Chen}
\affiliation{Zhejiang Key Laboratory of Micro-Nano Quantum Chips and Quantum Control, School of Physics, Zhejiang University, Hangzhou 310027, China}
\author{Yu-Yu Zhang}%
\email{yuyuzh@cqu.edu.cn}
\affiliation{Department of Physics, Chongqing Key Laboratory for strongly coupled Physics,
Chongqing University, Chongqing 401330, China}
\date{\today}

\begin{abstract}
We present a Hermitian squeezing-Dicke model as a minimal closed-system setting for non-Hermitian many-body physics. Squeezing generates an effective parity-time ($\mathcal{PT}$) symmetry and exchanges gainy and lossy eigenmodes without dissipation. We identify an anomalous dynamical superradiant phase (DSRP) that exhibits by a mean-field superradiant order in the $\mathcal{PT}$ symmetry-breaking regime, accompanied by a $\mathbb{Z}_2$-symmetry-breaking ground-state quantum phase transition. Additionally, it undergoes a $\mathcal{PT}$-breaking dynamical phase transition at an exceptional point (EP). The excitation energy vanishes near both boundaries, giving critical slowing down. The vacuum photon number is divergent at the ground-state transition but is finite and nonanalytic at the EP, distinguishing the DSRP from the conventional superradiant phase and from a standard dynamical instability.  Squeezing causes the coupling phase to act as an artificial magnetic flux, which yields flux-controlled nonreciprocal quantum amplification in the DSRP. Our results offer an alternative pathway to non-Hermitian many-body physics and nonreciprocal quantum amplification in closed Hermitian systems.

\end{abstract}

\maketitle

\emph{Introduction --}Light-matter coupling has brought
forth a novel class of quantum many-body systems by enabling photon-mediated interactions with ultracold atomic gas~\cite{naturephys2006,science1126,nature472}. A prominent example is the Dicke superradiant phase transition that has been realized in an cavity coupled to a Bose-Einstein condensate~\cite{PhysRev.93.99,nature2010,science2021,PhysRevA.78.051801}, which has a wide applications in quantum-enhanced metrology~\cite{PhysRevLett.122.113603,scinecelaser}. Recent advances in strongly coupled light-atom systems with synthetic magnetic fields include observations of chiral superradiant
phases~\cite{PhysRevLett.127.063602,PhysRevLett.129.183602}, frustrated superradiant phases~\cite{PhysRevLett.128.163601,PhysRevResearch.5.L042016,qj5x-t71k}, and fractional
quantum Hall physics~\cite{PhysRevA.93.023828, PhysRevLett.108.223602}. This progress motivates exploration of exotic superradiant phenomenon using external fields.

Recently, great interest has arisen in combining magnetic fields with broken Hermiticity to realize unconventional phases of matter. Non-Hermiticity with $\mathcal{PT}$ symmetry breaking leads to dynamical phase transitions associated with complex spectral singularities such as exceptional points~\cite{science2019,natremater2019}. Intriguing phenomenon have been illustrated in dissipation-induced no-Hermitian effects, including nonreciprocal phase transitions~\cite{PhysRevLett.131.113602,NATUREnonrec2021,PhysRevLett.132.193602}, as well as chiral dynamics in a dissipative Dicke model~\cite{PhysRevLett.122.193605,sciencechiral}. Indeed, such dissipative systems with engineered gain and loss provide a natural route to $\mathcal{PT}$ symmetry breaking. Recently, great interest has arisen in combining squeezing bosonic interactions with non-Hermiticity to explore novel chiral phononics, nonreciprocity~\cite{NATURE2022,NATURPHYS2023} and quantum amplification~\cite{Naturephys2021,PhysRevLett.114.093602}
By contrast, exploring genuinely exotic phases of matter with broken Hermiticity within a closed Hermitian framework remains a significant challenge.

Here, we propose a Hermitian squeezing-Dicke model accompanied by an effective $\mathcal{PT}$ symmetry breaking, differs from previous dissipation-based mechanisms~\cite{PhysRevLett.131.113602}. We identify a dynamical superradiant phase (DSRP) where a mean-field superradiant order coexists with a complex excitation spectrum. Unlike a generic dynamical instability, it retains the nonzero order parameter and criticality while exhibiting non-Hermitian dynamics. Our work establishes a genuinely novel closed-system phenomenon featuring effective non-Hermiticity, which exhibits both of a $\mathbb{Z}_2$-broken superradiant phase transition and a $\mathcal{PT}$-breaking dynamical phase transition. And we find nonreciprocity in the Hermitian system.

\begin{figure}   
\centering
\includegraphics[width=\linewidth]{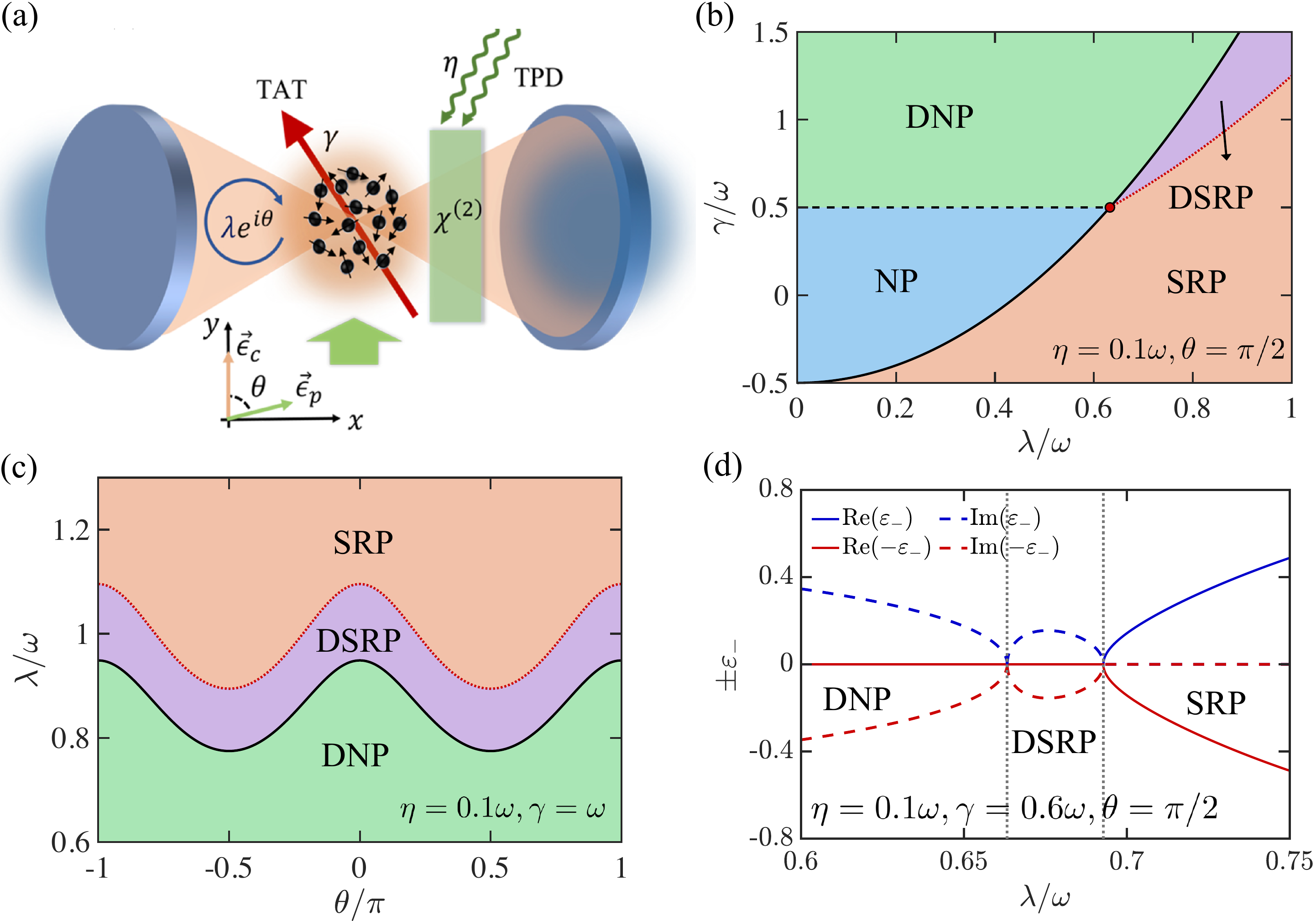}
\caption{(a) An ensemble of $N$ two-level atoms couples to a driven cavity with complex coupling $\lambda e^{i\theta}$. The green rectangle denotes the nonlinear crystal for two-photon driving. Ground-state phase diagrams in the $\lambda-\gamma$ (b) and $\theta-\lambda$ (c) planes.
The NP and DNP from the SRP and DSRP regions are separated by three critical boundaries
$\gamma_c$ (black dashed), $\lambda_c^{\mathrm{NP}}$ (black solid), and $\lambda_c^{\mathrm{SR}}$ (red dotted).
The red circle in (b) marks the multicritical point.} (d) BdG eigenvalues $\pm\varepsilon_-$ versus $\lambda$ across the
DNP $\rightarrow$ DSRP $\rightarrow$ SRP transition, exhibiting complex values in the DNP and DSRP regimes.
\label{fig1}
\end{figure}

\emph{Hamiltonian and effective $\mathcal{PT}$ symmetry--}We consider $N$ two-level atoms coupled to a single cavity mode through a linearly polarized laser inducing  complex light-atom couplings in Fig.~\ref{fig1}(a). The Hamiltonian is described by the Dicke model $\hat{H}_{\text{Dicke}} = \omega \hat{a}^{\dagger}\hat{a} + \frac{\lambda}{\sqrt{N}} \left(e^{-i\theta}\hat{a} + e^{i\theta}\hat{a}^\dagger\right) \left(\hat{J}_+ +\hat{J}_-\right) + \Delta\hat{J}_z$, 
where $\hat{a}\,\left(\hat{a}^{\dagger}\right)$ is the bosonic annihilation (creation) operator of the quantized cavity mode with frequency $\omega$, and $\Delta$ is a atom energy difference. The complex light-atom coupling is  $\lambda e^{i\theta}$, where $\theta$ is the angle between the laser and cavity polarization vectors~\cite{PhysRevLett.122.193605}.  The operator $\hat{J}_i = \sum_n \sigma_{i}^{n} \,\left(i = +,-,z\right)$ is a collective spin operator with  Pauli matrices $\sigma_i$. The Hamiltonian possesses a discrete $\mathbb{Z}_2$ symmetry for even and odd excitation parity.

Typically, we consider spins with ferromagnetic ($\gamma<0$) and antiferromagnetic ($\gamma>0$) mutual interaction $\gamma$ in the $xy$-spin plane~\cite{PhysRevA.71.064101}: $\hat{H}_I=2\gamma/N\sum_{n<m}\left(\sigma^{x}_{n}\sigma^x_{m} -\sigma^{y}_{n}\sigma^y_{m}\right)=2\gamma/N\left(\hat{J}^{2}_{x}-\hat{J}^{2}_{ y}\right)$. Meanwhile, a nonlinear crystal embedded into the cavity generates two-photon driving through optical parametric amplification~\cite{PhysRevLett.125.203601}. Our system can be described by a squeezing Dicke Hamiltonian
\begin{equation}\label{hamilt0}
\begin{aligned}
\hat{H} = \hat{H}_{\text{Dicke}}+ \eta \left(\hat{a}^{\dagger 2} + \hat{a}^2 \right) + \frac{2\gamma}{N}\left(\hat{J}^{2}_{x} - \hat{J}^{2}_{ y}\right),
\end{aligned}
\end{equation}
where $\eta>0$ is the strength of two-photon squeezing interactions. Crucially, squeezing makes the coupling phase $\theta$ gauge invariant, acting as an artificial magnetic flux.

We perform the Holstein-Primakoff transformation using  $\hat{J}_{+} = \hat{b}^{
    \dagger} \sqrt{N - \hat{b}^\dagger \hat{b}}
    $, $\hat{J}_{-} =  \sqrt{N - \hat{b}^\dagger \hat{b}}\hat{b}$ and $\hat{J}_{z} = \hat{b}^{\dagger}\hat{b} - \frac{N}{2}$, which represents angular momentum operators in terms of bosonic operators $b$  and $b^{\dagger}$.  The Hamiltonian becomes $\hat{H}_{\text{NP}} = \omega \hat{a}^{\dagger}\hat{a} + \lambda\left(e^{-i\theta} \hat{a} \hat{b}^{\dagger}+ e^{i\theta} \hat{a}^{\dagger}\hat{b}\right)+\Delta\left(\hat{b}^\dagger \hat{b} - \frac{N}{2}\right)+ \hat{H}_q$, where the squeezing terms are $
   \hat{H}_q =\lambda\left(e^{-i\theta} \hat{a}\hat{b}+ e^{i\theta} \hat{a}^{\dagger} \hat{b}^{\dagger}\right)+ \eta \left(\hat{a}^{\dagger 2} +\hat{a}^2\right) + \gamma \left(\hat{b}^{\dagger2} + \hat{b}^{2}\right)$. 
The Hamiltonian is bilinear in the creation and annihilation operators and can be diagonalized using a Bogoliubov transformation. Within the operator $\hat{\psi}_{\text{NP}} = (\hat{a},\hat{b}, \hat{a}^\dagger, \hat{b}^\dagger)^{T}$, the Hamiltonian can be expressed as $\hat{H}_\text{NP} = \frac{1}{2}\hat{\psi}_{\text{NP}}^{\dagger}\mathcal{M}_\text{NP}\hat{\psi}_{\text{NP}}$, which satisfies the bosonic commutation relation $[\hat{\psi}_{\text{NP}},\hat{\psi}_{\text{NP}}^{\dagger}]=\tau_z$  with $\tau_z = \sigma_{z}\otimes\mathit{I}$. The corresponding  Bogoliubov-de Gennes (BdG) Hamiltonian is given by $\mathcal{H}_{\text{BdG}}^\text{NP} = \tau_{z} \mathcal{M}_\text{NP}$ 
\begin{equation}\label{Bdg}
    \mathcal{H}_{\text{BdG}}^\text{NP} = \left(\begin{matrix}
        \omega & \lambda e^{i\theta} &2\eta &\lambda e^{i\theta}\\
        \lambda e^{-i\theta} &\Delta &\lambda e^{i\theta}&2\gamma\\
        -2\eta &-\lambda e^{-i\theta} &-\omega &-\lambda e^{-i\theta}\\
        -\lambda e^{-i\theta} &-2\gamma &-\lambda e^{i\theta} &-\Delta
    \end{matrix}
    \right).
\end{equation}
The quadrature interaction terms $H_q$ makes the BdG matrix $ \mathcal{H}_{\text{BdG}}^\text{NP}$ non-Hermitian. However, Hermiticity of $\hat{H}$ implies  $\tau_z\mathcal{H}_\text{BdG} \tau_z^{-1} = \mathcal{H}^{\dagger}_\text{BdG}$, showing the pseudo-Hermiticity symmetry of  $\mathcal{H}_{\text{BdG}}$.

The excitation spectrum of the original Hamiltonian (\ref{hamilt0}) is obtained by diagonalizing the BdG Hamiltonian
\begin{equation}
 \varepsilon_{\pm}^{\text{NP}} =\sqrt{\omega^2 - 2\left(\eta^2 + \gamma^2\right)\pm 2\sqrt{\epsilon}},
\end{equation}
where $\epsilon=\left(\gamma^2-\eta^2\right)^2+\lambda^2\left(\omega - 2\gamma\right)\left(\omega-2\eta\cos2\theta\right)$.
The vanishing of $\varepsilon_{-}^\text{NP}$ gives the critical coupling strength
\begin{equation}\label{lambda_c}
    \lambda_{c}^\text{NP} = \frac{1}{2}\sqrt{\frac{\left(\omega+2\gamma\right)\left(\omega^2-4\eta^2\right)}{\omega - 2\eta\cos2\theta}}.
\end{equation}
which is valid with the constraint conditions $-\omega/2<\eta\leq \omega/2$ and $-\omega/2\leq \gamma$. For weak coupling below $\lambda_{c}^\text{NP}$,  the ground state is in a normal phase (NP), in which the cavity field is empty and all spins point downward. When $\gamma\geq\gamma_c=\omega/2$, the real excitation
eigenenergy $\varepsilon_-^{\text{NP}}$ vanishes and the spectrum acquires an imaginary part, $\rm Im(\varepsilon_-^{\text{NP}})$, signaling a dynamical normal phase (DNP). In the DNP, the cavity remains empty but the excitations become dynamically unstable. $\lambda_{c}^\text{NP}$ is the phase boundary of the NP and the DNP, where $|\varepsilon_{-}^\text{NP}|$ vanishes in Fig.~\ref{fig1}(d).

The distinctive trait of the squeezing Dicke model is a modified $\mathcal{PT}$ symmetry even for nonzero flux $\theta$. One can identify suitable bases in which an inversion plane separating gain and loss on either side is clearly recognizable. We perform a unitary transformation $\mathcal{R}$ to diagonalize $H_q$ term (see SM), in which an inversion plane clearly separates gain and loss on opposite sides. For example, when $\theta=\pi/2$ and $\eta=-\gamma$, we analytically derives the effective gainy eigenmodes $\hat{X}_1 = \hat{X}_a - \hat{X}_b$, $\hat{X}_2 = \hat{Y}_a + \hat{Y}_b $ and the lossy modes $\hat{Y}_1 = i(\hat{Y}_a - \hat{Y}_b)$, $\hat{Y}_2 = i(\hat{X}_a + \hat{X}_b)$ , where the quadratures are given by $\hat{X}_{a} =  \frac{1}{2}(i\hat{a} - \hat{a}^{\dagger})$, $\hat{Y}_{a} = \frac{1}{2}(\hat{a} - i\hat{a}^{\dagger})$, $\hat{X}_{b} = \frac{1}{2}(\hat{b} - i\hat{b}^{\dagger})$, and $\hat{Y}_{b} = \frac{1}{2}(\hat{b}^{\dagger} - i\hat{b})$. In the quadrature basis $\{\hat{X}_1,\hat{X}_2,\hat{Y}_1,\hat{Y}_2\}$, the effective gain-loss Hamiltonian $\tilde{\mathcal{H}}_{\text{BdG}}^\text{NP}=\mathcal{R}^{\dagger}\mathcal{H}_\text{BdG}^\text{NP}\mathcal{R}$ is expressed as 
\begin{equation}\label{PTHAM}
\tilde{\mathcal{H}}_{\text{BdG}}^\text{NP} = \left(\begin{matrix}
i\xi_+ &  i\chi & \chi_+  & 0  \\
 -i\chi & i\xi_- &0   & \chi_-  \\
  \chi_+&0   & -i\xi_+ &  -i\chi \\
  0& \chi_-  & i\chi  & -i\xi_- 
\end{matrix}
\right),
\end{equation}
where the parameters are given by $\chi=(\omega-\Delta)/2$ and $\chi_{\pm}=\lambda\pm(\omega+\Delta)/2$. The gain rates for two modes $\hat{X}_1$ and $\hat{X}_2$ are  $\xi_{\pm} \equiv \lambda\pm2\gamma$, respectively.

The above Hamiltonian reveals an effective gain-loss mirror plane with the time-reversal $\mathcal{T}$ as complex conjugation by $i\leftrightarrow -i$ and a parity operation $\mathcal{P}_{XY}$ by exchanging $\hat{X}_i \leftrightarrow \hat{Y}_i$: $(\hat{X}_1, \hat{X}_2, \hat{Y}_1, \hat{Y}_2) \leftrightarrow (\hat{Y}_1, \hat{Y}_2, \hat{X}_1, \hat{X}_2)$. The BdG matrix $\tilde{\mathcal{H}}_{\text{BdG}}^\text{NP}$ can be mapped into a matrix invariant under $\mathcal{P}_{XY}\mathcal{T}$ operation, i.e. $\mathcal{P_{XY}T}\tilde{\mathcal{H}}_{\text{BdG}}^\text{NP}(\mathcal{P_{XY}T})^{-1}=\tilde{\mathcal{H}}_{\text{BdG}}^\text{NP}$ (see SM), exhibiting the effective $\mathcal{P}_{XY}\mathcal{T}$ symmetry.  Note that nonzero magnetic flux retain the effective $\mathcal{P}_{XY}\mathcal{T}$ symmetry, while it breaks the mirror symmetry $\mathcal {P}_{ab}$ by exchanging $\hat{a} \leftrightarrow\hat{b}$. Thus, the effective $\mathcal{PT}$-symmetric physics arise from squeezing interactions, in contrast to previous mechanisms with dissipation~\cite{PhysRevLett.131.113602}.

The eigenvalues of Eq.(\ref{PTHAM}) for resonance $\omega=\Delta$ are given by $(\pm\varepsilon_+,\pm\varepsilon_-)$ with excitation energies $\varepsilon_{\pm}=\sqrt{(\omega-2\eta\pm 2\lambda)(\omega+2\eta)}$. The lower energy $\varepsilon_-$ is real for $\lambda\leq\lambda_c^{\text{NP}}=\omega/2-\eta$ and $-\omega/2<\gamma<\omega/2$. When $\gamma>\omega/2$, the $\mathcal{P}_{XY}\mathcal{T}$ symmetry breaks with a complex $\varepsilon_-$, indicating the DNP governed by an effective non-Hermitian dynamics.

\emph{Non-Hermiticity of a superradiant phase--}For strong coupling $\lambda>\lambda_{c}^{\mathrm{NP}}$, photons in the cavity become macroscopically populated, and the system enters the superradiant regime. The bosonic operators are expected to shift as $\hat{a}\rightarrow \hat{a}+\sqrt{N}\alpha$ and $\hat{b}\rightarrow \hat{b} + \sqrt{N}\beta$, where the displacements are complex, $\alpha = A + iB$ and $\beta = X+iY$. In the thermodynamics limit $N\rightarrow\infty$, the Hamiltonian of Eq. (\ref{hamilt0}) becomes (see SM)
\begin{eqnarray}\label{hamilt3}
    \hat{H}_{\mathrm{SR}} &=& \omega \hat{a}^{\dagger}\hat{a} + \lambda^{\prime}\left(e^{-i\theta} \hat{a} + e^{i\theta} \hat{a}^{\dagger}\right)\left(\hat{b} + \hat{b}^\dagger\right) \nonumber\\
    &+&\Delta^{\prime}\hat{b}^\dagger \hat{b}  + \eta (\hat{a}^2+\hat{a}^{\dagger2} ) + \gamma^{\prime}(\hat{b}^2 +\hat{b}^{\dagger2} ).
\end{eqnarray}
The renormalized parameters are
$\lambda^{\prime}
=\lambda(1-2\beta^2)/\tilde{\beta}$,
$\Delta^{\prime}
=\Delta
-2\lambda\tilde{\alpha}/\tilde{\beta}
\allowbreak
-6\gamma\beta^2
\allowbreak
+\gamma\beta^4/\tilde{\beta}^2$,
and
$\gamma^{\prime}
=-\lambda\tilde{\alpha}/(2\tilde{\beta})
\allowbreak
+\gamma[
1-3\beta^2
\allowbreak
+\beta^4/(2\tilde{\beta}^2)]$,
where
$\tilde{\alpha}
=\beta(
e^{-i\theta}\alpha
\allowbreak
+e^{i\theta}\alpha^*)$,
$\tilde{\beta}
=\sqrt{1-\beta^2}$,
and $\beta$ is real. The linear terms in the $\hat{H}_\mathrm{SR} $ that are linear in the bosonic operators can be eliminated by choosing the displacement $\alpha$ and $\beta$, which are determined by mininizing the ground-state energy. By keeping only the terms proportional to $N$, we obtain the scaled ground-state energy
$E_g=\omega \left|\alpha\right|^2 + \Delta\left|\beta\right|^2 + \lambda \tilde{\beta}\left(e^{-i\theta}\alpha + e^{i\theta}\alpha^*\right)\left(\beta+\beta^*\right) +\gamma\tilde{\beta}^2\left(\beta^2 + \beta^{*2}\right)+ \eta\left(\alpha^2 + \alpha^{*2}\right)$. Minimizing $ E_g$ gives a complex $\alpha$ with $A =\pm\lambda\cos{\theta}\sqrt{\omega\left(\omega^2 - 4\eta^2\right)u_1+2u_1u_2}/[2\left(\omega + 2\eta\right)u_2]$ and $B=\pm\lambda\sin{\theta}\sqrt{\omega\left(\omega^2 - 4\eta^2\right)u_1+2u_1u_2}/[2\left(\omega - 2\eta\right)u_2]$, and a real $\beta$ with $X =\mp\sqrt{u_1/u_2}/2$ and $Y=0$,
where $u_1=4\lambda^2\left(\omega-2\eta\cos{2\theta}\right) - \left(\omega + 2\gamma\right)\left(\omega^2-4\eta^2\right)$ and $u_2 = 2\lambda^2\left(\omega-2\eta\cos{2\theta}\right)-\gamma\left(\omega^2-4\eta^2\right)$. 
Setting $\alpha=\beta=0$ yields the same critical coupling strength $\lambda_{c}^{\mathrm{NP}}= \frac{1}{2}\sqrt{\left(\omega+2\gamma\right)\left(\omega^2-4\eta^2\right)/(\omega-2\eta\cos2\theta)}$ in Eq.(\ref{lambda_c}).

\begin{figure} 
\centering
	\includegraphics[width=\linewidth]{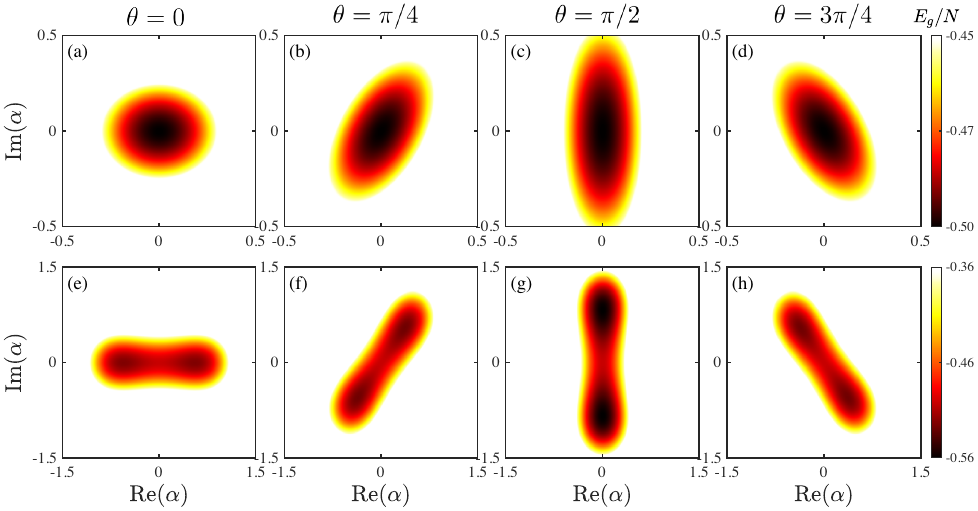}
\caption{
Scaled ground-state energy $E_g/N$ in the complex $\alpha$ plane for the DNP at $(\gamma,\lambda)/\omega=(1,0.7)$ [(a)–(d)] and for the DSRP at $(\gamma,\lambda)/\omega=(1,0.9)$ [(e)–(h)], with $\theta=0,\pi/4,\pi/2,$ and $3\pi/4$ and $\eta=0.1\omega$.
}
\label{fig3}
\end{figure}

Following the same procedure used in the NP, we uncover the effective $\mathcal{PT}$ symmetry in the superradiant regime. The bilinear Hamiltonian $\hat{H}_{\mathrm{SR}} $ (~\ref{hamilt3}) can be written as $\hat{H}_\mathrm{SR} = \frac{1}{2}\hat{\psi}_{\mathrm{SR}}^{\dagger}\mathcal{M}_\mathrm{SR}\hat{\psi}_{\mathrm{SR}}$ in terms of $\hat{\psi}_{\mathrm{SR}} = (\hat{a},\hat{b}, \hat{a}^\dagger, \hat{b}^\dagger)^{T}$. The non-Hermitian BdG Hamiltonian is obtained as $\mathcal{H}_{\text{BdG}}^\mathrm{SR} = \tau_{z} \mathcal{M}_\mathrm{SR}$. We dervie the effective gian-loss Hamiltonian $\tilde{\mathcal{H}}_{\text{BdG}}^\text{SR}=\mathcal{R}^{\dagger}\mathcal{H}_\text{BdG}^\text{SR}\mathcal{R}$ in terms of gainy and lossy eigenmodes, which respects effective $\mathcal{P}_{XY}\mathcal{T}$ symmetry (see SM). 
Two pairs of eigenvalues of $\mathcal{H}_{\text{BdG}}^\mathrm{SR}$ are obtained as $(\pm\varepsilon_{+}^{\mathrm{SR}},\pm\varepsilon_-^{\mathrm{SR}})$, where $\varepsilon_{\pm}^{\mathrm{SR}}$ are excitation energies,
\begin{equation}\label{excitedenergy}
    \varepsilon_{\pm}^{\mathrm{SR}} =\sqrt{\omega^2 + \Delta^{\prime2} - 4\left(\eta^2 + \gamma^{\prime2}\right)\pm \sqrt{\epsilon^{\prime}}}/\sqrt{2}, 
\end{equation}
with $\epsilon^{\prime}=(\omega^2+4\gamma^{\prime2}-\Delta^{\prime2}-4\eta^2)^2+16\lambda^{\prime2}(\Delta^{\prime} - 2\gamma^{\prime})(\omega-2\eta\cos2\theta)$. 
The lower eigenenergy $\varepsilon_{-}^{\mathrm{SR}} $ becomes complex, when $\omega^2 + \Delta^{\prime2} - 4\left(\eta^2 + \gamma^{\prime2}\right)-\sqrt{\epsilon^{\prime}}<0$. The eigenenergies $\pm\varepsilon_{-}^{\mathrm{SR}}$ coalesce at an exceptional point (EP) $\lambda_{c}^{\mathrm{SR}}$. At the EP, $|\varepsilon_{-}^{\mathrm{SR}}|$ vanishes, analogous to the gap closing at the critical point $\lambda_{c}^{\mathrm{NP}}$, which yields the critical coupling of the EP
\begin{equation}
\label{lambda_el}
\lambda_{c}^{\mathrm{SR}} = \sqrt{\frac{\gamma\left(\omega^2-4\eta^2\right)}{\omega-2\eta\cos 2\theta}}.
\end{equation}
Fig.~\ref{fig1}(d) shows complex eigenenergies
with $\rm Im(\pm\varepsilon_-^{\mathrm{SR}})$ for $\lambda_{c}^\text{NP}<\lambda<\lambda_{c}^\text{SR}$, in contrast to real values when $\lambda>\lambda_{c}^\text{SR}$. 

Fig.~\ref{fig1}(b) shows the ground-state phase diagram. The DSRP with complex excitation energy $\varepsilon_{-}^{\mathrm{SR}}$ is separated from the SRP with real eigenenergies by $\lambda_{c}^\text{SR}$.
Notably, the DSRP arises as an intermediate phase between the DNP and
SRP in the regime $\lambda_{c}^{\mathrm{NP}}<\lambda<\lambda_{c}^{\mathrm{SR}}$  for arbitrary $\theta$
in Fig.~\ref{fig1}(c). The order
parameter $\alpha$ reveals a second-order transition from the DNP to the DSRP in Fig.~\ref{fig3}.
In the DNP, $\lambda<\lambda_{c}^{\mathrm{NP}}$,
the mean-field energy landscape has a single minimum at
$\alpha=0$. In contrast, the DSRP displays two degenerate minima, signaling spontaneous $\mathbb{Z}_2$ symmetry breaking. As $\theta$ increases from $0$ to $3\pi/4$, the pair of minima rotates anticlockwise in the complex plane in Fig.~\ref{fig3}(e)--(h), which demonstrates that $\theta$ acts as a synthetic magnetic flux controlling the orientation of the energy landscape.

\begin{figure}
    \centering
	\includegraphics[width=\linewidth]{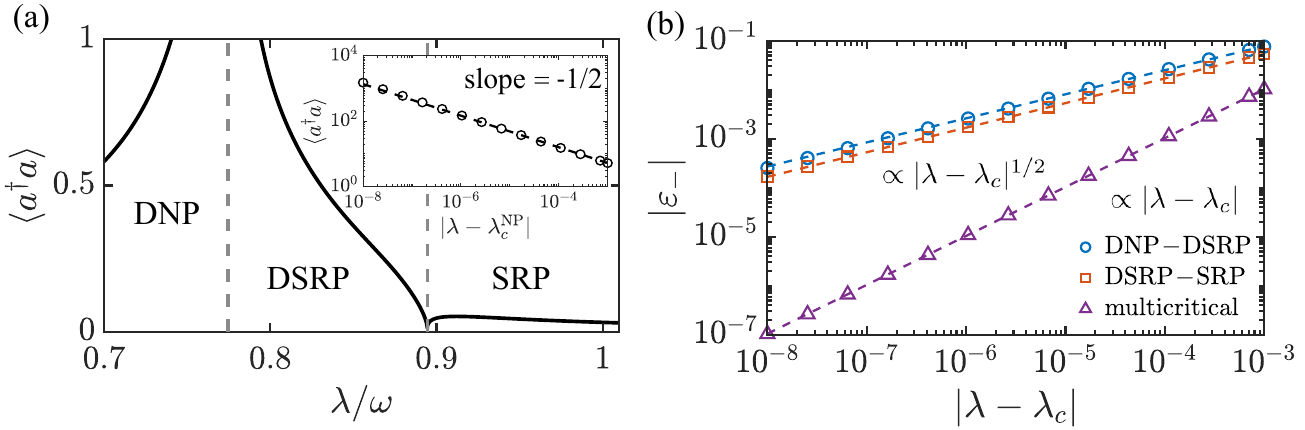}
    \caption{(a) Averge photon number $\langle a^\dagger a\rangle$ in the Bogoliubov vacuum state as a function of $\lambda$, with the inset showing the scaling relation near the DNP--DSRP boundary. (b) Lowest excitation energy $|\varepsilon_-|$ as a function of $|\lambda-\lambda_c|$ on a log-log scale.}
    \label{scaling}
\end{figure}

Besides breaking $\mathbb{Z}_2$ symmetry the DSRP breaks $\mathcal{PT}$ symmetry, whereas the SRP restores the $\mathcal{PT}$ symmetry above the EP. The imaginary part of  $\varepsilon_{-}^{\mathrm{SR}} $ induces exponential mode growth, unlike the bounded oscillations of the SRP. Beyond the complex energies, the DSRP and SRP can also be distinguished by the average photon number $\langle a^\dagger a\rangle$ in the Bogoliubov vacuum state $\mathcal{R}|0\rangle$, which characterizes the quantum fluctuations near criticality. It diverges at the critical value $\lambda_c^{\mathrm{NP}}$ as $|\lambda-\lambda_c^{\mathrm{NP}}|^{-1/2}$, while it remains finite at the EP $\lambda_c^{\mathrm{SR}}$ in Fig.~\ref{scaling}(a). This nonanalyticity physically predicts a dynamical transition from the DSRP to SRP at the EP, which is distinct from the DNP-DSRP ground-state transition.

As $\lambda$ approaches the critical value $\lambda_c^{\mathrm{NP}}$, the excitation energy in both DNP and DSRP vanishes as  $|\varepsilon_-|\propto|\lambda-\lambda_c|^{1/2}$ , while the multicritical point exhibits a distinct scaling $|\varepsilon_-|\propto|\lambda-\lambda_{\mathrm{mc}}|$, as shown in Fig.~\ref{scaling}(b). Similarly, the excitation gap closes with the same $1/2$ exponent near the EP $\lambda_c^{\mathrm{SR}}$ for the DSRP--SRP transition. The closing of the gap is accompanied by critical slowing down, with a characteristic relaxation time $\tau_{\mathrm{cr}}\sim|\varepsilon_-|^{-1}$. Accordingly, $\tau_{\mathrm{cr}}\propto|\lambda-\lambda_c^{\mathrm{NP}}|^{-1/2}$ at both DNP-DSRP and DSRP-SRP transitions, indicating the divergence of the relaxation time near these critical points.

\emph{Nonreciprocal amplification in the DSRP--}We  investigate dynamical signatures to differentiate the DSRP from the SRP. The dynamics in the superradiant regimes is governed by the non-Hermitian BdG effective Hamiltonian $\mathcal{H}_{\text{BdG}}^\mathrm{SR}$. The operator $\hat{\psi} = (\hat{a},\hat{b}, \hat{a}^\dagger, \hat{b}^\dagger)^{T}$ obeys the Heisenberg equation of motion $id\hat{\psi}/dt=-[\hat{H}_{\mathrm{SR}},\hat{\psi}]=\mathcal{H}_{\text{BdG}}\hat{\psi}$,
where $\mathcal{H}_{\text{BdG}}= \tau_{z} \hat{H}_{\mathrm{SR}}$ is the non-Hermitian BdG matrix. Assume $\left|\phi_n \right\rangle$ be an eigenvector of  $\mathcal{H}_\text{BdG}$ with eigenvalue $\varepsilon_n$.  Then $\tau_{z}\left| \phi_n \right\rangle$ is an eigenvector of  $\mathcal{H}^{\dagger}_\text{BdG}$ with the same eigenvalue $\varepsilon_n$ due to the pseudo-Hermiticity symmetry $\tau_z\mathcal{H}_\text{BdG} \tau_z^{-1} = \mathcal{H}^{\dagger}_\text{BdG}$. Two sets of biorthogonal eigenvectors $|\phi_n\rangle$ and $|\varphi_n\rangle$ corresponding to eigenvalues $\varepsilon_n$ and $\varepsilon_m^*$ satisfy a $\tau_z$-orthonormal relation: $\langle\varphi_m|\tau_z|\phi_n\rangle = \delta_{mn}$~\cite{PhysRevLett.132.220402,PhysRevE.57.6101,Fu2025,NATURE2022}. The BdG Hamiltonian can be expressed in terms of the completeness relation $\mathcal{H}_\text{BdG}=\sum_n\varepsilon_n|\phi_n\rangle \left\langle\varphi_{n} \right|\tau_{z}$. The time-dynamics operator $\hat{O}\left(t\right)$ is governed by $ \hat{O}\left(t\right)=\exp \left( -i\mathcal{H}_\text{BdG}t\right)\hat{O}\left(0\right)$, where an arbitrary initial vector is $\hat{O}\left(0\right) =\left\langle v^{0} \right| \hat{\psi}$. For example, $\hat{O}\left(0\right)=\hat{a}$ when $\langle v^0| = \left(1, 0, 0, 0\right)$. We project $ \hat{O}\left(t\right)$ onto $\tau_z$-orthogonal basis, yielding
$\hat{O}\left(t\right) = \sum_{n} e^{-i\varepsilon_n t} \langle v^{0}|\phi_{n}\rangle\langle\varphi_{n}| \tau_z\hat{\psi}$.

\begin{figure}
    \centering
	\includegraphics[width=\linewidth]{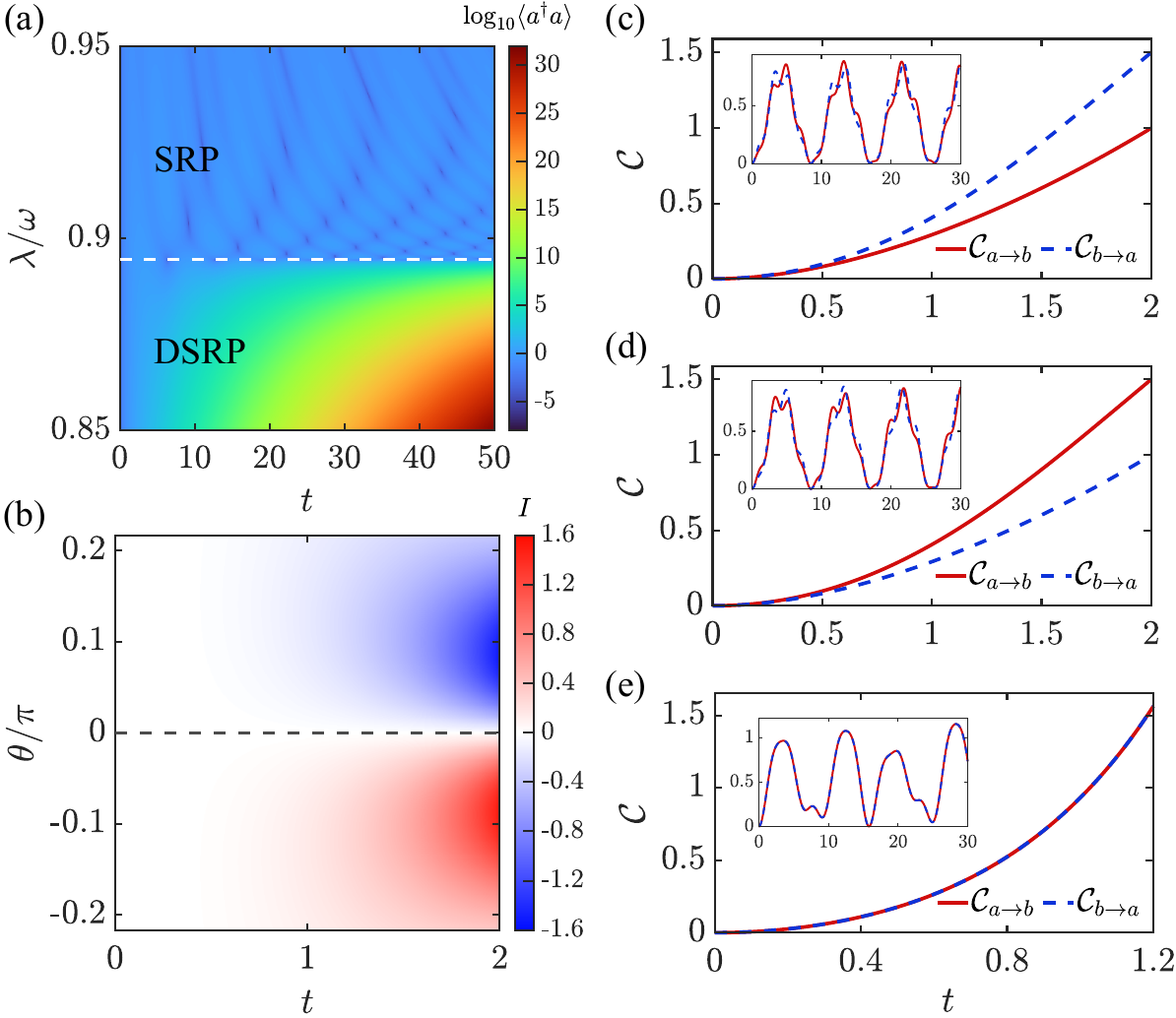}
    \caption{(a) Dynamics of the photon number $\langle a^\dagger a(t)\rangle$ in the DSRP and SRP. The white dashed line marks the EP $\lambda_c^{\mathrm{SR}}$. (b) Nonreciprocal transmission imbalance $I(t)$ in the DSRP under flux control by $\theta$. The transmission ratios of DSRP $\mathcal{C}_{a\to b}$ (red solid) and $\mathcal{C}_{b\to a}$ (blue dashed) are shown for $\theta=0.2\pi$ (c), $-0.2\pi$ (d), and $0$ (e), respectively. The insets in (c)--(e) show the corresponding transmission ratios in the SRP. Other parameters are fixed at $\eta=0.1\omega$ and $\gamma=\omega$.
    }
\label{fig5}
\end{figure}

Fig.~\ref{fig5}(a) illustrates the dynamics of the average photon number $\langle a^\dagger(t) a(t)\rangle$ as a function of the coupling strength. In the DSRP, the photon number  exhibits exponential growth induced by the complex eigenvalues $\varepsilon_n=\pm\varepsilon_{-}^{\mathrm{SR}}$ for $\lambda<\lambda_c^{\mathrm{SR}}$. Conversely, the SRP, exhibiting bounded periodic dynamics for $\lambda>\lambda_c^{\mathrm{SR}}$. This transition across the EP signifies the dynamical phase transition from the SRP to the DSRP at the EP in a stability boundary, which is induced by spontaneous breaking of the effective $\mathcal{PT}$ symmetry.

To explore the nonreciprocal dynamics, we explore the transsion between atoms and photons. When the system is initially driven in the photonic subsystem with a single photon and no atomic excitation, i.e. the initial state $|10\rangle$, the excitations flow to the atomic subsystem is $n_{b}(t)=\langle10|\hat{b}^\dagger \left(t\right)\hat{b}(t)|10\rangle$. The transmission efficiency of excitations transfer from photons to atoms is given by the ratio $\mathcal{C}_{{a}\rightarrow {b}} = n_{b}(t)/{n}_{a}\left(0\right)$ (see SM). When the atom subsystem is driven with an empty cavity, i.e. the initial state $|01\rangle$, the instantaneous photon number is given by $n_{a}(t)=\langle01|\hat{a}^\dagger \left(t\right)\hat{a}(t)|01\rangle$. It leads to the atom-to-photon transmission ratio $\mathcal{C}_{{b}\rightarrow {a}} = n_{a}(t)/n_{b}\left(0\right)$. To quantify the nonreciprocal, we introduce transmission imbalance $I(t)=\mathcal{C}_{a\rightarrow b}(t)-\mathcal{C}_{b\rightarrow a}(t)$,
where its sign identifies the nonreciprocal direction.

Fig.~\ref{fig5}(b) shows the flux-controlled nonreciprocal transmission imbalance $I(t)$ in the DSRP. For $\theta <0$, an initial excitation in photon mode $a$ is amplified as it is transferred to the atomic mod $b$ , while being rapidly attenuated in the opposite
direction. This Nonreciprocal amplification arises from the $\mathcal{PT}$ breaking, enabling quantum amplification.
For zero flux, $\theta=0$, initial excitation either bosonic modes leads to reciprocal dynamics,
$\mathcal{C}_{a\rightarrow b}=\mathcal{C}_{b\rightarrow a}$,
as shown in Fig.~\ref{fig5}(e).
For $\theta=0.2\pi$, Fig.~\ref{fig5}(c) shows that an initial excitation in the atomic mode $b$ is amplified and transferred to the photonic mode $a$,
yielding $\mathcal{C}_{b\rightarrow a}>1$, while the reverse process
$\mathcal{C}_{a\rightarrow b}$ is suppressed.
Reversing the flux to $\theta=-0.2\pi$ reverses the preferred direction, favoring amplified transfer from the photonic mode $a$ to the atomic mode $b$.
The resulting $\mathcal{C}_{a\rightarrow b}\neq\mathcal{C}_{b\rightarrow a}$ demonstrates flux-tunable nonreciprocal quantum amplification in the DSRP, with the direction controlled by the sign of $\theta$.
The insets of Figs.~\ref{fig5} show the bounded periodic dynamics in the SRP.

\emph{Conclusion--}We find non-Hermiticity emerges in a Hermitian squeezing-Dicke model generating a genuine stationary superradiant phase. The DSRP is anomalous in combining superradiant order with broken effective $\mathcal{PT}$ symmetry. The vacuum photon number is divergent at the ground-state transition but is finite and nonanalytic at the EP, distinguishing the DSRP from the SRP. This nonanalyticity and the critical slowing down at the EP distinguish the DSRP from a standard instability. These features establish the DSRP as a distinct phase bounded by a $\mathbb{Z}_2$-symmetry-breaking quantum phase transition and a $\mathcal{PT}$-breaking dynamical transition at the EP. Moreover, unlike the SRP, which exhibits bounded oscillations, the DSRP exhibits flux-controlled nonreciprocal quantum amplification. These results establish a closed Hermitian route to non-Hermitian many-body physics and nonreciprocal quantum devices.

\textit{Acknowledgments--}Y.-Y. Z. acknowledges support from the National Natural Science Foundation of China Grant No.12475013, No.12547101, and the Fundamental ResearchFunds for the Central Universities Grant No. 2025CDJ-IAISYB-029. Q. H. C. acknowledge support from the National Key R$\&$D Program of China under Grant No. 2024YFA1408900.

\bibliography{apssamp}%
\end{document}